\documentclass[twocolumn,letterpaper,aps,prd,superscriptaddress,floatfix]{revtex4}

\usepackage{graphicx}	

\usepackage{xspace}	

\begin{document}

\title{ Erratum:  Measurement of transverse single-spin asymmetries 
for $J/\psi$ production in polarized p+p collisions 
at $\sqrt{s}$ = 200 GeV [Phys. Rev. D 82, 112008 (2010)] }

\author{A.~Adare}  
\author{S.~Afanasiev}  
\author{C.~Aidala}  
\author{N.N.~Ajitanand}  
\author{Y.~Akiba}  
\author{H.~Al-Bataineh}  
\author{J.~Alexander}  
\author{H.~Al-Ta'ani}  
\author{A.~Angerami}  
\author{K.~Aoki}  
% ...
\collaboration{PHENIX Collaboration} \noaffiliation



\maketitle

\begin{figure}[hbt]
\includegraphics[width=1.0\linewidth]{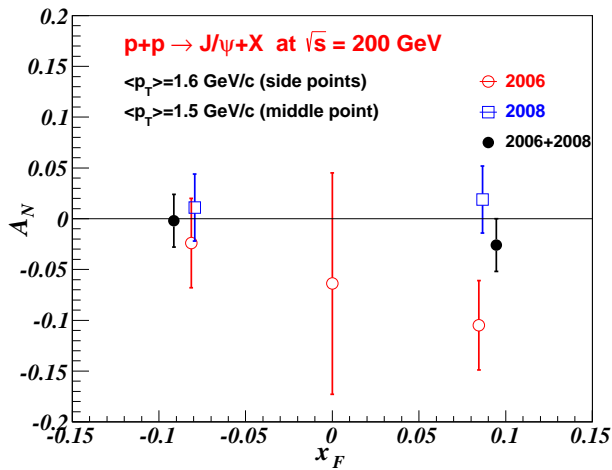}
\caption{\label{fig:NewFig3}  (color online) 
Transverse single-spin asymmetry in $J/\psi$ production as a 
function of $x_F$ for 2006 and 2008 data sets separately, and the 
combined result; the points for the combined result have been 
offset by 0.01 in $x_F$ for visibility. The error bars shown are 
statistical and type A systematic uncertainties, added in 
quadrature. Type B systematic uncertainties are not included but 
are 0.003 or less in absolute magnitude and can be found in 
Table~\ref{tab:AN_erratum}. Not shown is an additional uncertainty 
in the scale of the ordinate due to correlated polarization 
uncertainties of 3.4\%, 3.0\%, and 2.4\% for the 2006, 2008, and 
combined 2006 + 2008 data sets, respectively.}
\end{figure}

\begin{figure}[hbt]
\includegraphics[width=1.0\linewidth]{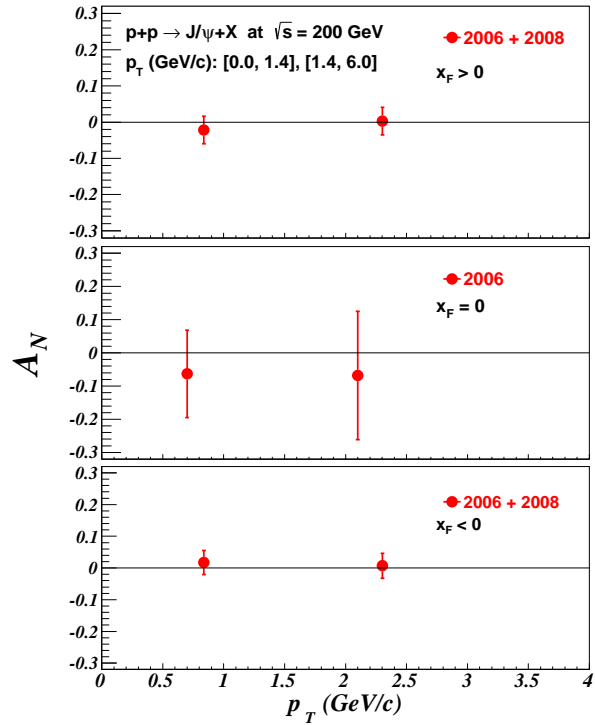}
\caption{\label{fig:NewFig4} (color online) 
Transverse single-spin asymmetry of $J/\psi$ mesons plotted against 
$J/\psi$ transverse momentum. See Table~\ref{tab:AN_erratum} for 
mean $x_F$ values for each point. The error bars shown are 
statistical and type A systematic uncertainties, added in 
quadrature. Type B systematic uncertainties are not included but 
are 0.002 or less in absolute magnitude and can be found in 
Table~\ref{tab:AN_erratum}. An additional uncertainty in the scale 
of the ordinate due to correlated polarization uncertainties is 
2.4\% (3.4\%) for the points with $|x_f| > 0$. ($x_F = 0$ is not 
shown.)}
\end{figure}

We previously reported~\cite{ppg103} measurements of transverse 
single-spin asymmetries, $A_N$, in $J/\psi$ production from transversely 
polarized $p+p$ collisions at $\sqrt{s} = 200$~GeV with data taken by 
the PHENIX experiment at the Relativistic Heavy Ion Collider in 2006 and 
2008.  Subsequently, we have found errors in the analysis procedures for 
the 2008 data, which resulted in an erroneous value for the extracted 
$A_N$.  The errors affected the sorting of events into the correct 
left/right and forward/backward bins.  This produced an incorrect value 
for the 2008 result, but the 2006 result is unaffected.  We have 
conducted two independent reanalyses with these errors corrected, and we 
present here the corrected values for the 2008 data and the combined 
results for 2006 and 2008.  Figures~\ref{fig:NewFig3} and 
\ref{fig:NewFig4} replace Figs. 3 and 4 in~\cite{ppg103}.  
Tables~\ref{tab:Bg-asymmetries_erratum}, \ref{tab:Bg-fraction_erratum}, 
and \ref{tab:AN_erratum} replace Tables I, II, and V, respectively, 
in~\cite{ppg103}.

We also made a small change in the way $\left<x_F\right>$ 
is calculated; this did not affect the asymmetries.  In 
the original paper, we calculated $\left<x_F\right>$ 
separately for the North and South muon spectrometers, 
but this does not exactly represent the way the data are 
combined to form an asymmetry; we fixed this in 
Tables~\ref{tab:Bg-asymmetries_erratum}--\ref{tab:AN_erratum}.
The new combined spin asymmetry in the forward region is 
$A_N = -0.026 \pm 0.026 {\rm (stat)} \pm 0.003 {\rm 
(sys)}$.  Since this asymmetry is consistent with zero, we 
no longer claim that our results suggest a possible non-zero 
trigluon correlation function in transversely polarized protons.


\clearpage

\begin{table*}[th]
\begin{minipage}[t]{0.48\linewidth} \centering
\caption{\label{tab:Bg-asymmetries_erratum}  
Background asymmetries as a function of $p_{T}$ for PHENIX muon 
spectrometers.  The uncertainties given are statistical.}
\begin{ruledtabular}\begin{tabular}{cccc}
$p_T$ (GeV/$c$)&  $\left<x_{F}\right>$  &   data set   &  $A_N^{BG}$  \\ \hline
0--6           &   -0.083     &     2006    & -0.003$\pm$0.028  \\
               &   -0.084     &     2008    & 0.041$\pm$0.035  \\
               &    0.083     &     2006    & -0.008$\pm$0.028  \\
               &    0.084     &     2008    & -0.010$\pm$0.036  \\
\\
0--1.4         &   -0.083     &     2006    & -0.021$\pm$0.034  \\
               &   -0.085     &     2008    & 0.033$\pm$0.043  \\
               &    0.083     &     2006    &  0.002$\pm$0.034  \\
               &    0.085     &     2008    & 0.023$\pm$0.044  \\
\\
1.4--6         &   -0.083     &     2006    &  0.001$\pm$0.053  \\
               &   -0.084     &     2008    & 0.019$\pm$ 0.075 \\
               &    0.083     &     2006    & -0.039$\pm$0.053  \\
               &    0.084     &     2008    &  -0.038$\pm$0.076  \\
\end{tabular}\end{ruledtabular}
\end{minipage}
\hspace{0.2cm}
\begin{minipage}[t]{0.48\linewidth} \centering
\caption{\label{tab:Bg-fraction_erratum}
Total background fractions as a function of $p_{T}$ for muon 
spectrometers on the north and south sides of PHENIX.  Backgrounds 
were higher in the 2006 data set because the less restrictive 
trigger requirement allowed more random track combinations.}
\begin{ruledtabular}\begin{tabular}{ccccc}
$p_T$ (GeV/$c$)   &    data set   & \ \ detector \ \  & background fraction (\%)\\ \hline
0--6              &     2006     &       South       & 21.7$\pm$0.6    \\
                  &     2006     &       North       & 19.1$\pm$0.4    \\
                  &     2008     &       South       & 16.2$\pm$0.2    \\
                  &     2008     &       North       & 14.1$\pm$0.2    \\
\\
0--1.4            &     2006     &       South       & 23.2$\pm$0.7    \\
                  &     2006     &       North       & 22.0$\pm$0.7    \\
                  &     2008     &       South       & 15.9$\pm$0.3    \\
                  &     2008     &       North       & 15.6$\pm$0.3    \\ 
\\
1.4--6            &     2006     &       South       & 20.1$\pm$0.8    \\
                  &     2006     &       North       & 14.1$\pm$0.5    \\
                  &     2008     &       South       & 15.7$\pm$0.4    \\
                  &     2008     &       North       & 9.7$\pm$0.2    \\ 
\end{tabular}\end{ruledtabular}
\end{minipage}

\vspace{0.5cm}

\caption{\label{tab:AN_erratum}
$A_N$ vs. $p_T$ in forward, backward and midrapidity.  Systematic 
uncertainties in the last two columns are due to the geometric 
scale factor and the polarization, respectively.  There are 
additional Type C uncertainties due to the polarization of 3.4\%, 
3.0\%, and 2.4\% for the 2006, 2008, and combined 2006 and 2008 
results.}
\begin{ruledtabular}\begin{tabular}{cccccccc}
 $p_T$      & Data Sample& $\left<x_F\right>$ & $A_N$& $\delta A_N$ & $\delta A_N$ & $\delta A_N^f$ (\%) & $\delta A_N^P$ (\%) \\
(GeV/c)     &            &          &        & (stat.)  & (Type A syst.)& (Type B syst.)      &  (Type B syst.)     \\ \hline
           & 2006        & -0.083   & -0.024  & 0.044   & 0.003         & 0.6                 & 2.3 \\
           & 2008        & -0.084   & 0.011  & 0.033   & 0.004         & 0.4                 & 3.4 \\
           & 2006 + 2008 & -0.084   & -0.002  &0.026   & 0.002         & 0.4                 & 2.8 \\ 
\\
0--6       & 2006        &  0.000   & -0.064  & 0.106   & 0.026         & 0.6                 & 2.3 \\
           & 2006        &  0.083   & -0.105  & 0.044   & 0.005         & 0.6                 & 2.3 \\
           & 2008        &  0.084   & 0.019  & 0.033   & 0.003         & 0.4                 & 3.3 \\
           & 2006 + 2008 &  0.084& -0.026&  0.026  & 0.003         & 0.4                 & 2.7 \\ 
\\           
            & 2006        & -0.083   &  0.050  & 0.067  & 0.007         & 0.6                 & 2.3 \\
            & 2008        & -0.085   & 0.001   & 0.047   & 0.008         & 0.4                 & 3.4 \\
            & 2006 + 2008 & -0.084   & 0.017  & 0.038  & 0.005         & 0.4                 & 2.8 \\ 
\\
0--1.4      & 2006        & 0.000    & -0.063  & 0.128  & 0.031         & 0.6                 & 2.3\\
            & 2006        & 0.083   & -0.065  & 0.066   & 0.005         & 0.6                 & 2.3 \\
            & 2008        &  0.085   & 0.0003  & 0.047   & 0.003         & 0.4                 & 3.4 \\
            & 2006 + 2008 &  0.084  & -0.022  & 0.038  & 0.003         & 0.4                 & 2.7 \\ 
\\          
            & 2006        & -0.083   & -0.073  & 0.065 & 0.002          & 0.6                 & 2.3 \\
            & 2008        & -0.084   & 0.051  & 0.048   & 0.010          & 0.4                 & 3.5 \\
            & 2006 + 2008 & -0.084   & 0.007  &0.039  & 0.002          & 0.4                 & 2.8 \\
\\
1.4--6      & 2006        & 0.000   & -0.068  & 0.188  & 0.045          & 1.2                 & 2.3 \\
\\
            & 2006        & 0.083   & -0.046  & 0.064  & 0.005          & 0.6                 & 2.3 \\
            & 2008        & 0.084    & 0.030  & 0.047   & 0.007          & 0.4                 & 3.3 \\
            & 2006 + 2008 & 0.084  & 0.003  &0.038  & 0.004          & 0.4                 & 2.7 \\
\end{tabular}\end{ruledtabular}
\end{table*}

\end{document}